\documentclass[aps,twocolumn,superscriptaddress, showpacs,floatfix]{revtex4}
\usepackage{graphicx}
\usepackage{dcolumn}
\usepackage{bm}

\def\esym{$E_{sym}(\rho)$~}

\def\rpi{$\pi^-/\pi^+$~}

\def\et{$\eta$~}

\begin{document}

\title{Effects of nuclear symmetry energy on $\eta$ meson production and its rare decay to the dark U-boson in heavy-ion reactions}

\author{Gao-Chan Yong}\email{yonggaochan@impcas.ac.cn}
\affiliation{Institute of Modern Physics, Chinese Academy of Sciences, Lanzhou 730000, China}
\author{Bao-An Li}\email{Bao-An.Li@tamuc.edu}
\affiliation{Department of Physics and Astronomy,
Texas A\&M University-Commerce, Commerce, TX 75429, USA\\
and Department of Applied Physics, Xi'an Jiaotong University, Xi'an 710049, China}

\begin{abstract}
Using a relativistic transport model ART1.0, we explore effects of nuclear symmetry energy on \et meson production and its rare decay to the dark U-boson
in heavy-ion reactions from 0.2 to 10 GeV/nucleon available at several current and future facilities. The yield of \et mesons at sub-threshold energies is found to be very
sensitive to the density dependence of nuclear symmetry energy. Above a beam energy of about 5 GeV/nucleon in Au+Au reactions, the sensitivity to symmetry energy disappears. Using the branching ratio of the rare \et decay ($\eta \rightarrow \gamma U$) available in the literature, we estimate the maximum cross section for the U-boson production in the energy range considered, providing a useful reference for future U-boson search using heavy-ion reactions.
\end{abstract}

\pacs{25.75.-q, 21.65.Mn, 21.65.Ef, 95.35.+d} \maketitle

\section{Introduction}
The physics motivations of this work are twofold. The first
purpose is to look for new and possibly more sensitive probes of
the high-density behavior of nuclear symmetry energy in high
energy heavy-ion collisions. The other purpose is to explore the
possibility of producing the neutral vector U-boson (dark photon)
introduced in the super-symmetric extension of the Standard Model
(SM) \cite{Boe-prl,Boe04,fayet06,bor06,fayet09} in relativistic
heavy-ion collisions. The two purposes naturally come together in
studying properties of neutron stars where both the possible
existence of the U-boson and the high-density behavior of nuclear
symmetry energy affect significantly the Equation of State (EOS)
of dense neutron-rich nucleonic matter. In fact, it has been shown
already that massive neutron stars can be supported even with a
rather soft EOS for symmetric nuclear matter and/or a super-soft
symmetry energy if the additional interaction between two nucleons
due to the exchange of a U-boson is considered \cite{Kri09,wen09}.
Moreover, both the symmetry energy and the possible U-boson affect
the core-crust transition density/pressure and the moment of
inertia of neutron stars \cite{zheng12}.

\subsection{Potentials of using \textbf{$\eta$} meson production in heavy-ion collisions as a probe of the high-density behavior of nuclear symmetry energy}
Nuclear symmetry energy \esym encodes the energy cost of
converting all protons into neutrons in symmetric nuclear matter.
Because the density dependence of \esym is very important for
understanding properties of rare isotopes and neutron stars as
well as the dynamics of heavy-ion reactions, supernova explosions
and gravitation wave emissions of spiraling neutron star binaries,
much efforts have been devoted to extracting the \esym using data
from both terrestrial laboratory experiments and astrophysical
observations, see e.g., refs.
\cite{LiBA98,Bro00,li2,Lat04,Bar05,Ste05a,LCK08,eos2011,newton12,Chen2012,Li2012,Lattimer12,Tsang12,Far13}.
While significant progress has been made recently in constraining
the $E_{\rm sym}(\rho_0)$ and the slope $L(\rho_0) \equiv \left[3
\rho (\partial E_{\rm sym}/\partial \rho\right]_{\rho_0}$  at
normal density $\rho_0$, see, e.g., refs.
\cite{LWChen11,Moller,Pawel09,agr12,sun10,tsa1,tsa2,chen05a,li05a,shet,Jim,Cen09,War09,Chen10,Mliu10,XuLiChen10a,Kli07,Car10,mye96,Koh10,Vid12,Wen12,Steiner10,Steiner12,Mike,Sot12,Newton09},
our knowledge about the high-density behavior of the \esym is
still very poor. In fact, findings about the high-density \esym
from comparing heavy-ion reaction data with various transport
model calculations are still inconsistent, see, e.g., refs.
\cite{xiao09,feng10,russ11,cozma11}. This is at least partially
because of both the uncertainties in the physics inputs and the
different numerical techniques in initializing and modeling the
transport process of colliding nuclei, see, e.g., refs.
\cite{zhyx08,dds11,mdp,cozma11,epj,init} for discussions on some
of the contributing factors. Thus, new and possibly more sensitive
probes of the high-density behavior of nuclear symmetry energy are
always useful. Since the symmetry (isovector) potential is
normally much smaller than the isoscalar potential, in order to
observe clearly effects of the symmetry energy/potential in
heavy-ion reactions one needs ideally to use slowly moving
particles to be acted on by the isovector potential for a long
time in a large region of high isospin asymmetry and density
gradients. Sub-threshold mesons, such as pions and kaons, especially
the ratios of different charge states, are good candidates and have
been studied extensively. Transport model
calculations have shown consistently that near their respective
production threshold, the \rpi \cite{LiBA02,Gai04,LiBA05a,Fer05p,zhang09}
and $K^+/K^0$ \cite{Fer05k} ratios are indeed sensitive to the high-density
symmetry energy although comparisons with limited data available
are still inconclusive so far. Generally speaking, more massive
mesons probe the \esym at higher densities if effects of the final
state interactions do not wash out the signal. The \et meson of
mass 547.853 MeV/c$^2$ \cite{eta08} is the most massive member in
the octet of pseudoscalar Goldstone mesons. It has significant
photon and dilepton decay channels providing the possibility of
studying the high-density symmetry energy more cleanly using the
electromagnetic probes. In fact, \et meson production in heavy-ion
collisions has been studied extensively both theoretically and
experimentally, especially by the TAPS and HADES Collaborations,
see, e.g., refs.
\cite{dep89,mosel91,wolf93,metag93,berg94,taps00,cassing01,Aga10,Aga12,Sal13}.
Moreover, as pointed out earlier \cite{dep89,mosel91}, because of
the hidden strangeness (the $s\bar{s}$ component), \et mesons
experience weaker final state interactions compared to pions.
Contrary to kaons, because the net strangeness content is zero in
\et mesons, they can be produced without another strange partner
in the final state and thus require less energies. More
quantitatively, the threshold invariant energy for \et and kaon
production in nucleon-nucleon collisions is
$(\sqrt{s})_{th}^{\eta}=2.4$ GeV and $(\sqrt{s})_{th}^{K}=2.6$
GeV, respectively. The corresponding threshold beam energy in free
nucleon-nucleon collisions is $E_{th}^{\eta}=1.2$ GeV and
$E_{th}^{K}=1.6$ GeV, respectively. For a comparison, the
experimental $K^+/\pi^+$ and $\eta/\pi^0$ ratios in the most
central Au+Au reactions at 1 GeV/nucleon is approximately $2\times
10^{-3}$ \cite{senger00} and $1.4\times 10^{-2}$ \cite{Ave03},
respectively. It is thus interesting to know if the \et meson
might be useful for exploring the \esym. Intuitively, in order for
the p-n pair to create a \et near its production threshold, they
have to have their momenta oriented opposite to each other. Thus,
near threshold \et mesons will probe sensitively the p-n relative
momentum which is determined mainly by the gradient of the
isovector potential. Compared to pions and kaons, however, the
elementary \et production cross sections in baryon-baryon and
meson-baryon scattering still suffer from relatively larger
uncertainties although they are gradually better known as more
data and calculations become available \cite{wolf93,Tho07}. It is
thus necessary to know the optimal beam energies and the kinematic
regions where the yield of \et mesons is most sensitive to the
\esym.

\subsection{Heavy-ion reactions as a possible tool for dark U-boson search}
While there are well established observational evidences for dark
matter through its gravitational interactions, the mass and
interactions of dark matter particles are completely unknown
\cite{kan08}. Thus, what is dark matter? It is actually at the
very top of the Eleven Science Questions for the New Century
identified by the Committee on the Physics of the Universe, US
National Research Council \cite{11questions}. In fact, much
efforts have been devoted to searching for dark matter candidates,
see, e.g., refs. \cite{jungman96,bertone05} for reviews. Besides
the weakly interacting massive particles as the most popular
candidates for dark matter, it has been proposed that light dark
matter particles $\chi$ with a mass in the range of 0.5-20 MeV may
exist \cite{Boe-prl,Boe04,fayet06,bor06,fayet09}. In particular,
the annihilation of $\chi \bar \chi\rightarrow e^+e^-$ through the
exchange of the light vector U-boson of mass 10-100 MeV, has been
used to explain successfully the INTEGRAL satellite observation of
the excess flux of 511 keV photons from the center region of our
galaxy \cite{511kev} although alternative explanations exist
\cite{Wei08}. Moreover, assuming the U-boson couples to quarks as
well as electrons \cite{kan08}, the extra contribution to the pion
decay ($\pi^0\rightarrow e^+e^-$) width mediated by an off-shell
U-boson can explain nicely the enhanced pion decay observed by the
KTeV Collaboration \cite{KteV} in comparison with the Standard
Model prediction \cite{dor07}. Furthermore, as mentioned earlier,
the extra interaction between nucleons due to the U-boson exchange
may affect properties of neutron stars
\cite{Kri09,wen09,zheng12,Zhang11} although it has no effect on
properties of finite nuclei \cite{Xu13}. It may also cause the
deviation from the inverse-square-law of gravity, see, e.g.,
\cite{Adel03,Fis99,Rey05,New09,Uzan03,Astro2010} for recent
reviews. The aforementioned studies and findings have motivated
considerable interest in searching for the U-boson \cite{reece09}.
While various upper limits on the strength and interaction range
of the U-boson have been put forward at various length scales
without contradicting known physics principles and existing
experimental/observational data, see, e.g., ref. \cite{Xu13} for
the latest review, most of the constraints on properties of the
U-boson are indirect. It is thus very interesting to see that some
direct constraints on this kind of gauge bosons through neutral
meson decays in neutrino experiments at CERN have been reported recently
\cite{Gninenko1201}. Moreover, several proposals to
investigate directly properties of the U-boson in low energy
($E_{c.m}\leq$ 10 GeV) experiments have been put forward. These
include experiments using $e^+e^-$ colliders through the
$e^{+}e^{-}\rightarrow \gamma U$ process
\cite{reece09,barze2011,bor06,zhu07,li10}, meson rare decays at
meson factories and fixed target electron-nucleus scatterings
\cite{reece09}. Many mesons can decay into the
U-boson with branching ratio $BR (X \rightarrow Y+U) \approx
\epsilon^2 BR (X \rightarrow Y + \gamma)$ where $\epsilon$ is the strength
ratio of the U-SM particle to $\gamma$-SM particle coupling.
This is then followed by the U decay into dileptons, i.e., $U \to \ell^+ \ell^-$.
Using the meson summary tables of the Particle Data Group, Reece and
Wang estimated the U-producing branching ratios for several mesons
\cite{reece09}. The \et rare decay was shown to be the most
promising one. More quantitatively, the ratio of $\eta \rightarrow
\gamma U$, $\omega \rightarrow \pi^0 U$, $\phi \rightarrow \eta U$
and $K_L^0 \rightarrow \gamma U$ is found to be approximately
$320/225/1.3/4.4\times 10^{-3}$. It is interesting to note that
the $\gamma+p\rightarrow \eta+p$ reaction with a cross section of
approximately 70 nb will be used for the U-boson search at the
Jlab Eta Factory \cite{LGan}. As we shall estimate in this work,
the \et production cross section in a Au+Au collisions from 0.2 to
10 GeV/A beam energy range is on the order of 60-6000 mb. While it
will be very challenging to extract information about the U-boson
from various backgrounds, the relatively large \et production
cross section warrants a U-boson search using heavy-ion
facilities, such as the CSR (Cooler Storage Ring) and the planned
HIAF (High Current Heavy-Ion Accelerator Facility) in Lanzhou, the
FAIR (Facility for Antiproton and Ion Research) in Darmstadt, NICA
(Nuclotron based Ion Collider Facility) at JINR Dubna and the BES
(Beam Energy Scan) program at RHIC (Relativistic Heavy-Ion
Collider). To explore the idea more quantitatively, we shall
investigate the excitation function of \et production in Au+Au
reactions in the energy range covered by these facilities.

\section{Transport model simulations of \textbf{$\eta$} meson production in heavy-ion collisions}
\begin{figure}[th]
\begin{center}
\includegraphics[width=0.45\textwidth]{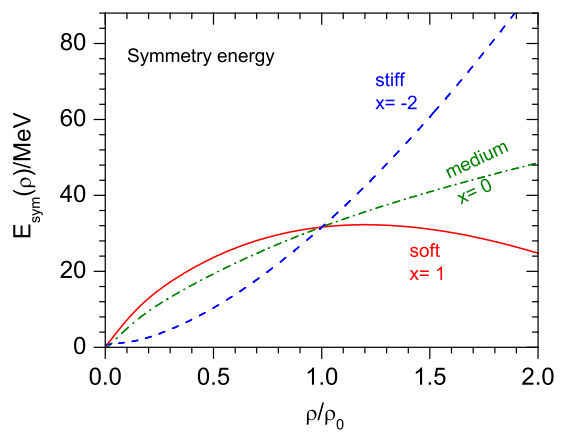}
\end{center}
\caption{(Color online) Nuclear symmetry energy as a function of reduced density $\rho/\rho_0$ (the symmetry energy at $\rho_0$ is taken to be 31.6 MeV) with three values for the parameter $x$. Taken from Ref.~\cite{chen05}.} \label{esym}
\end{figure}

Our study is based on the relativistic transport model ART
\cite{li95}. The model was developed from the BUU transport model
\cite{bertsch} by including more baryon and meson resonances and
their interactions. More specifically, the following baryons
$N,~\Delta(1232),~N^{*}(1440),~N^{*}(1535),~\Lambda,~\Sigma$, and
mesons $\pi,~\rho,~\omega,~\eta,~K$ with their explicit isospin
degrees of freedom are included. The model was used successfully
in studying many features of heavy-ion reactions at AGS energies
up to a beam momentum of about 15 GeV/c, for a review, see, ref.
\cite{ARTreview}. An extended version of ART is also used as a
hadronic afterburner in the AMPT (A Multi-Phase Transport Model)
model for heavy-ion collisions at RHIC and LHC energies
\cite{ampt}. In the ART model, $\eta$ meson production and
absorption are modeled via $\eta N \leftrightarrow N^*(1535)$. The
cross sections for the production and absorption of $N^{*}(1535)$
resonance in baryon-baryon collisions and its decay width can be
found in refs. \cite{wolf93,li95}. For ease of the following
discussions, we emphasize that $\sigma(NN\rightarrow
NN^{*}(1535))\approx 2\sigma(NN\rightarrow NN\eta)$ and the
elementary cross section for \et production is strongly isospin
dependent, i.e., $\sigma(pn\rightarrow pn\eta)\approx
3\sigma(pp\rightarrow pp\eta)=3\sigma(nn\rightarrow nn\eta)$. The
ART1.0 has the option of using different isoscalar potentials for
baryons. However, the isovector (symmetry) potential was not
implemented earlier. For this exploratory study, we use the
Skyrme-type parametrization for the isospin-dependent but
momentum-independent mean field potential \cite{li95,chen05}
\begin{eqnarray}
U(\rho ,\delta ,\tau )&\equiv&A(\rho /\rho _{0})+B(\rho
/\rho _{0})^{\sigma} \nonumber\\
&+&4\tau E_{\text{sym}}^{\mathrm{pot}}(\rho )+(18.6-F(x)) \nonumber\\
&\times&(G(x)-1)(\rho /\rho _{0})^{G(x)}\delta ^{2}. \label{Umf}
\end{eqnarray}
In the above, $E_{sym}^{pot}(\rho) = F(x)\rho /\rho_{0}
+(18.6-F(x))(\rho /\rho _{0})^{G(x)}$) is the interaction part of
nuclear symmetry energy, $\delta\equiv(\rho_{n}-\rho_{p})/\rho$ is the isospin asymmetry at
density $\rho$ and $\tau$=$1/2 (-1/2)$ for neutrons (protons).
Adding the kinetic part of nuclear symmetry energy $E_{sym}^{kin}(\rho) = \hbar^{2}/6m_{n}(3\pi^{2}\rho/2)^{2/3}$ where $m_n$ is the nucleon mass,
the symmetry energy corresponding to the single-particle potential in Eq. \ref{Umf} is
\begin{eqnarray}\label{esymF}
E_{sym}(\rho)&=& E_{sym}^{kin}(\rho)+ E_{sym}^{pot}(\rho)\nonumber\\
&=&\hbar^{2}/6m_{n}(3\pi^{2}\rho/2)^{2/3}\nonumber\\
&+& F(x)\rho /\rho_{0} +(18.6-F(x))(\rho /\rho _{0})^{G(x)}
\end{eqnarray}
with $F(x)$ and $G(x)$ given in Ref.~\cite{chen05} for different values of the parameter x used to vary the density dependence of $E_{sym}$ \cite{MDI}. Shown in Fig. \ref{esym} is the \esym
with $x= 1, 0$ and $-2$, respectively, from being very soft to very stiff. The mean-field potentials of mesons and baryon resonances in nuclear matter are still largely unknown.
We adopt here the minimum assumption that the isoscalar part of the mean-field potential for baryon resonances is the
same as that for nucleons and there is no mean-field for mesons as assumed in most transport models.
The isovector potential for baryon resonances is taken as an average of that for neutrons and protons with an isospin-dependent weighting factor
determined by the square of the Clebsch-Gordan coefficients in the $\Delta (N^*)\leftrightarrow \pi+N$ coupling \cite{LiBA02}.
Contributions of the different charge states of baryon resonances to the isospin asymmetry of the baryonic system are similarly calculated \cite{LiBA02}.
We choose two sets of parameters A and B, i.e., A= -356 (-124) MeV,
B= 303 (70.5) MeV and $\sigma$= 7/6 (2) corresponding to the
incompressibility of nuclear matter K$\sim$ 200 (380) MeV at $\rho_{0}=0.168~fm^{-3}$.

\section{Effects of nuclear symmetry energy on \textbf{$\eta$} meson production}

\begin{figure}[th]
\begin{center}
\includegraphics[width=0.45\textwidth]{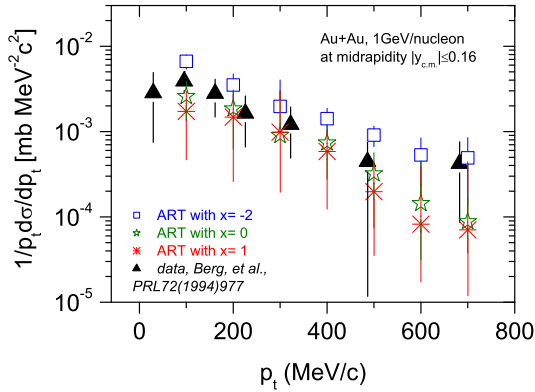}
\end{center}
\caption{(Color online) Transverse momentum distributions of
midrapidity $\eta$ mesons in inclusive Au+Au reaction at a
beam energy of 1 GeV/nucleon with different nuclear symmetry
energies. Data are taken from Ref.~\cite{berg94}.} \label{data94}
\end{figure}
Shown in Fig.~\ref{data94} is a comparison of the ART1.0 model calculations with the TAPS experimental data on the \et transverse momentum distribution at midrapidity
in Au+Au reaction at a beam energy of 1 GeV/nucleon. In this and the following model calculations, the error bars are statistical in nature. It is seen that our calculations give a reasonable description of the data with $x=0$ and 1 while the stiff nuclear symmetry energy with x= -2 leads to a significantly larger number of $\eta$ mesons. For a comparison, we notice that the FOPI $\pi^-/\pi^+$ data in Au+Au reactions at beam energies from about 0.4 to 1 GeV/nucleon favor the symmetry energy with x=1 \cite{xiao09}. Because \et is sensitive to the number of p-n collisions while the $\pi^-/\pi^+$ ratio is determined by the ratio of n-n and p-p colliding pairs, \et and the $\pi^-/\pi^+$ ratio provide complementary information about the symmetry energy. It has been argued that sub-threshold \et production requires multiple nucleon-nucleon scattering over sufficiently long time \cite{dep89}, it may thus carry information mainly about the equilibrium phase of the reaction. At this stage, the following condition is satisfied between any two regions of density (isospin asymmetry) $\rho_1(\delta_1)$ and $\rho_2(\delta_2)$ \cite{shi00,li02npa}
\begin{eqnarray}
E_{sym}(\rho_{1})\delta_{1} = E_{sym}(\rho_{2})\delta_{2}.
\end{eqnarray}
Then, the so-called isospin fractionation occurs, i.e., the density region with a higher symmetry energy $E_{sym}(\rho)$ will have a lower isospin asymmetry $\delta$ and vice versa.
Consequently, with a stiffer symmetry energy the isospin asymmetry $\delta$ is lower in the high density regions where then more $np~(pn)$ collisions can occur to produce more $\eta$ mesons.
While it is the opposite for the $\pi^-/\pi^+$ ratio.

\begin{figure}[th]
\begin{center}
\includegraphics[width=0.45\textwidth]{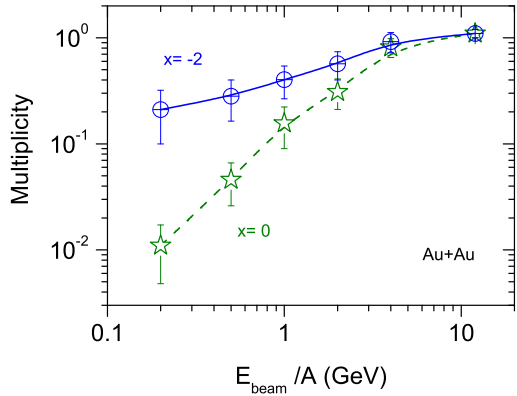}
\end{center}
\caption{(Color online) Multiplicity of inclusive \et production
as a function of incident beam energy in Au+Au reactions with the
two different values for the symmetry energy parameter x.} \label{exc}
\end{figure}
It is well known that the $\pi^-/\pi^+$ ratio is more sensitive to
the symmetry energy at lower beam energies especially in the
sub-threshold region where the mean-field dominates the reaction
dynamics and has longer time to modify the momentum of nucleons.
To create a \et at sub-threshold energies requires even longer
reaction time for both multiple collisions and the mean-field to
act coherently in order to accumulate enough energy on the two colliding
baryons or their resonances. Thus, it is interesting to examine
the multiplicity of \et production as a function of incident beam
energy. Shown in Fig.~\ref{exc} are the \et multiplicity as a
function of beam energy in inclusive Au+Au reactions with both
$x=0$ and $x=-2$. It is seen that the $\eta$ multiplicity
decreases rapidly with decreasing beam energy, especially for the
soft symmetry energy. The $\eta$ multiplicity saturates at an
incident energy of about 10 GeV/nucleon. It is interesting to see
that indeed the effect of nuclear symmetry energy is stronger in
the deeper sub-threshold region as one expects. In fact, it is
more sensitive to the symmetry energy than the $\pi^-/\pi^+$ ratio
in the same energy region ~\cite{zhf12} although the pion yields
are relatively more abundant.
\begin{figure}[th]
\begin{center}
\includegraphics[width=0.45\textwidth]{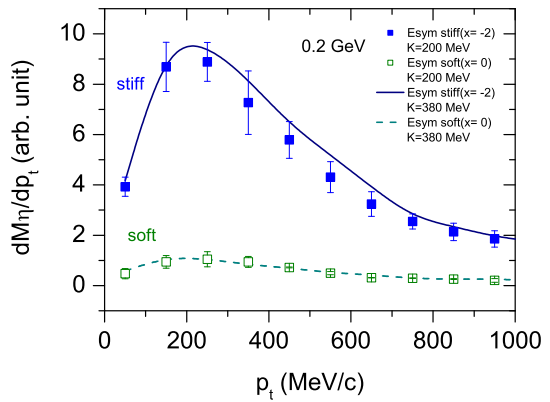}
\end{center}
\caption{(Color online) Transverse momentum distributions of
$\eta$ production in inclusive Au+Au reaction at incident beam
energy of 0.2 GeV/nucleon with different nuclear symmetry energies
and compression coefficients of symmetric matter.} \label{tran}
\end{figure}
More detailed information about \et production and its dependence
on the symmetry energy can be obtained from studying its
transverse momentum distribution. Shown in Fig.~\ref{tran} is such a
distribution in terms of $dM_{\eta}/dp_{t}\equiv
\frac{dM_{\eta}(p_{t} \rightarrow p_{t}+dp_{t})}{dp_{t}}$ for Au+Au reactions at a beam energy of 0.2
GeV/nucleon with different values for the parameter x and the
incompressibility $K$. It is seen that at a transverse momentum of about $200 \sim 300$
MeV/c, the effect of nuclear symmetry energy reaches its maxima of
about a factor of 5-10. Changing the incompressibility of nuclear
matter from K=200 MeV to K=380 MeV seems to have little effect on
the \et yields and its spectrum. While the larger (smaller)
incompressibility of nuclear matter causes smaller (larger)
compression during the reaction, both the production and
reabsorption of \et are approximately equally affected. Moreover,
as mentioned earlier, the p-n relative momentum is essentially
determined by the gradient of the isovector potential with little
influence from the isoscalar potential. Thus, the near-threshold \et
production is sensitive to the symmetry energy but not the
incompressibility of symmetric nuclear matter.

\section{Production of dark U-boson from the rare decay of \textbf{$\eta$} mesons in heavy-ion collisions}
\begin{figure}[th]
\centering
\includegraphics[width=0.45\textwidth]{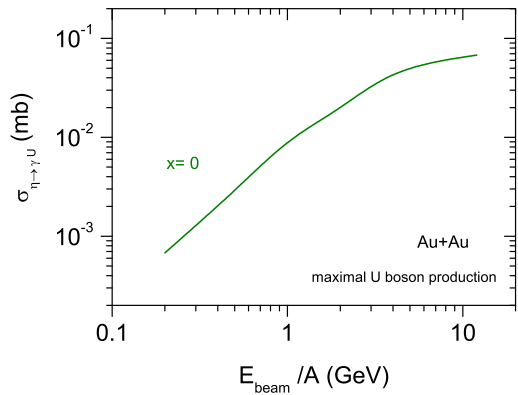}
\caption{(Color online) The maximal possible U-boson production cross section from the $\eta$ rare decay in inclusive Au+Au reaction
as a function of beam energy with the soft symmetry energy (x= 0).}
\label{U-boson}
\end{figure}
We now turn to estimating the maximum U-boson production cross
section. For this purpose, we use the maximal U-boson-SM particle
coupling constant $\epsilon^2 = 4.\times 10^{-6}$ and $BR (X
\rightarrow Y + \gamma) \simeq 0.7$ \cite{reece09,eta08}). Since
the symmetry energy with x = 0 can better reproduce the \et
production data from the TAPS collaboration, we present here our estimates using the soft symmetry energy with x = 0.
Using the stiff symmetry energy with x=-2 would increase the maximal
total U-boson production cross section by the same ratio shown in
Fig.~\ref{exc}. Shown in Fig.~\ref{U-boson} is the maximal total
U-boson production cross section in Au+Au reaction at a beam
energy from 0.2 to 10 GeV/nucleon. It is seen that the maximum
U-boson production cross section is in the 1-70 $\mu$b range which is
about 6-8 orders of magnitude higher than that estimated for the
$\gamma+p$ reaction at the Jlab Eta Factory. However, similar to
the situations at other facilities, it will be a big challenge to find
in the dileption spectrum in heavy-ion collisions the signature of
the dark U-boson decay $U \to \ell^+ \ell^-$ because of the
various backgrounds. Moreover, the predicted U-boson width is only about 10 eV which
is much smaller than the resolution of most detectors.
Nevertheless, our estimate of the excitation function of U-boson
production in heavy-ion collisions provide a useful reference for
future U-boson search.

\section{Conclusions}
In summary, within the transport model ART1.0 we studied the
excitation function of \et yield in heavy-ion collisions both as a
potential probe of the symmetry energy and as a possible source of the dark U-boson.
The yield of \et meson is found to be very sensitive to the
density dependence of nuclear symmetry energy at sub-threshold
energies. Using the branching ratio of the rare \et decay ($\eta
\rightarrow \gamma U$) available in the literature, we estimated
the maximum cross section for the U-boson production in heavy-ion
collisions at a beam energy of 0.2 to 10 GeV/nucleon, providing a useful reference for future U-boson search experimentally.

\section{Acknowledgements}
This work was supported in part by the National Natural Science
Foundation of China under Grant No. 11175219, the IMP principal
fund, the U.S. National Science Foundation grants PHY-0757839 and
PHY-1068022, the U.S. National Aeronautics and Space
Administration under grant NNX11AC41G issued through the Science
Mission Directorate, and through CUSTIPEN (China-U.S. Theory Institute
for Physics with Exotic Nuclei) under U.S. DOE grant number DE-FG02-13ER42025.

\end{document}